# The crossmetric tensor and the geometrical meaning of the imaginary numbers in quaternions

**Cyril Cayron**[a]

[a] Ecole Polytechnique Fédérale de Lausanne, EPFL-STI-LMTM, rue de la Maladière 70, 2000 Neuchâtel, Switzerland

**Synopsis** The composition rule of rotational crystallographic quaternions previously written as scalar and cross product is now translated as a quadratic product with a 4x4 matrix called symbolic crossmetric tensor. A geometrical meaning for the imaginary symbols of the Cartesian and crystallographic quaternions is proposed that is perfectly coherent with the quadratic form.

**Abstract** The product of two Cartesian quaternions can be written as a quadratic form based on a 4x4 matrix made of the imaginary numbers $\mathbf{i}, \mathbf{j}, \mathbf{k}$ introduced by Hamilton. We generalized this matrix to non-Cartesian bases and showed how to build it from the metric and the cross tensors. We called it "crossmetric tensor". Its symbols are the imaginary numbers $\mathbf{i}_a, \mathbf{i}_b, \mathbf{i}_c$ related to the crystallographic basis. The form of the crossmetric tensors was determined for the six crystal families. We also showed that the unit crystallographic quaternion can be geometrically represented by an infinity of pairs of oriented planes intersecting along the vectorial component of the quaternion such that the angle between them is the semiangle of the rotation. The composition of quaternions follows the source-target (groupoid) rule which says that the target plane of the first quaternion should coincide with the source plane of the second quaternion, and the result is the quaternion formed by the source plane of the former and the target plane of the latter. The elementary quaternions are based on the crystallographic vectors $\mathbf{a}, \mathbf{b}, \mathbf{c}$; they are geometrically represented by all the pairs of perpendicular oriented planes intersecting along the axes $\mathbf{a}, \mathbf{b}, \mathbf{c}$, respectively. Other "complementary" quaternions were also introduced as the pairs of planes formed among the three faces of the unit cell. The elementary crystallographic quaternions follow Hamilton's rules on the squares of imaginary numbers, and the complementary quaternions follow Hamilton's rules on the bi and tri-products. The Cartesian quaternions $\mathbf{i}, \mathbf{j}, \mathbf{k}$ are specific cases of crystallographic quaternions. They are formed by the pairs of faces of the cube. Since the faces intersect along the cube axes and are perpendicular to each other, the elementary and complementary quaternions are equal, which explains why the square, bi and tri-product Hamilton's rules are satisfied all together with only three quaternions $\mathbf{i}, \mathbf{j}, \mathbf{k}$ and not six.

## 1. Introduction

### 1.1. Reminder on the rotational crystallographic quaternions

In a previous work (Cayron, 2026), we generalized the way to write quaternions for non-Cartesian bases. For a rotation of angle α around the axis $\mathbf{u}$, the associated crystallographic quaternion is

$$\mathbf{q} = \mathrm{c} + \mathbf{u} \quad (1)$$

with $\mathbf{u} = s\, \tilde{\mathbf{u}}$, with c and s the cosine and sine of the semi-angle of the rotation, $\mathrm{c} = \cos(\frac{\alpha}{2})$ and $\mathrm{s} = \sin(\frac{\alpha}{2})$, and the tilde over $\mathbf{u}$ means that this vector is normalized in the crystallographic basis $\boldsymbol{\mathcal{B}}_c$, i.e. $||\tilde{\mathbf{u}}|| = \sqrt{\tilde{\mathbf{u}}^{\mathrm{t}}\, \boldsymbol{\mathcal{M}}\, \tilde{\mathbf{u}}} = 1$, with $\boldsymbol{\mathcal{M}}$ the metric tensor. We recall that

$$\boldsymbol{\mathcal{M}} = \begin{pmatrix} \mathbf{a}^2 & \mathbf{b}^{\mathrm{t}}\,\mathbf{a} & \mathbf{c}^{\mathrm{t}}\,\mathbf{a} \\ \mathbf{a}^{\mathrm{t}}\,\mathbf{b} & \mathbf{b}^2 & \mathbf{c}^{\mathrm{t}}\,\mathbf{b} \\ \mathbf{a}^{\mathrm{t}}\,\mathbf{c} & \mathbf{b}^{\mathrm{t}}\,\mathbf{c} & \mathbf{c}^2 \end{pmatrix} = \begin{pmatrix} ||\mathbf{a}||^2 & ||\mathbf{b}||\,||\mathbf{a}||\cos(\gamma) & ||\mathbf{c}||\,||\mathbf{a}||\cos(\beta) \\ ||\mathbf{a}||\,||\mathbf{b}||\cos(\gamma) & ||\mathbf{b}||^2 & ||\mathbf{c}||\,||\mathbf{b}||\cos(\alpha) \\ ||\mathbf{a}||\,||\mathbf{c}||\cos(\beta) & ||\mathbf{b}||\,||\mathbf{c}||\cos(\alpha) & ||\mathbf{c}||^2 \end{pmatrix} \quad (2)$$

The metric tensor is a coordinate transformation matrix from the reciprocal basis to the direct basis, $\boldsymbol{\mathcal{M}} = [\boldsymbol{\mathcal{B}}_c^* \rightarrow \boldsymbol{\mathcal{B}}_c]$ . It allows the calculation of the scalar products between vectors $\mathbf{u}$ and $\mathbf{v}$ by the quadratic form

$$\mathbf{u} \cdot \mathbf{v} = \mathbf{u}^{\mathrm{t}}\, \boldsymbol{\mathcal{M}}\, \mathbf{v} \quad (3)$$

If two rotations $\mathcal{R}_1$ and $\mathcal{R}_2$ are represented by the crystallographic quaternions $\mathbf{q}_1 = \mathrm{c}_1 + \mathbf{u}_1$ and $\mathbf{q}_2 = \mathrm{c}_2 + \mathbf{u}_2$, respectively, their composition is the crystallographic quaternion $\mathbf{q}_3$ obtained by generalizing the scalar product and the cross product such that

$$\mathbf{q}_3 = \mathbf{q}_1\, \mathbf{q}_2 = \mathrm{c}_3 + \mathbf{u}_3 \quad \text{with} \quad (4)$$

$$\mathrm{c}_3 = \mathrm{c}_1 \mathrm{c}_2 - (\mathbf{u}_1^{\mathrm{t}}\, \boldsymbol{\mathcal{M}}\, \mathbf{u}_2) \quad (5)$$

$$\mathbf{u}_3 = \mathrm{c}_2\, \mathbf{u}_1 + \mathrm{c}_1 \mathbf{u}_2 + \boldsymbol{\mathcal{X}}\left(\mathbf{u}_1 \underline{\times}\; \mathbf{u}_2\right) \quad (6)$$

with $\underline{\times}$ the usual Cartesian cross product, and $\boldsymbol{\mathcal{X}}$ the cross matrix directly deduced from the metric tensor $\boldsymbol{\mathcal{M}}$ by

$$\boldsymbol{\mathcal{X}} = \sqrt{\det(\boldsymbol{\mathcal{M}})}\, \boldsymbol{\mathcal{M}}^{-1} \quad (7)$$

Note that $\sqrt{\det(\boldsymbol{\mathcal{M}})} = V = \det(\boldsymbol{a}, \boldsymbol{b}, \boldsymbol{c})$, the volume of unit cell.

### 1.2. Hamilton quaternions and composition tensor for Cartesian bases

Quaternions were historically introduced for Cartesian bases in the form

$$\mathbf{q} = s + x\,\mathbf{i} + y\,\mathbf{j} + z\,\mathbf{k} \quad (8)$$

with $s, x, y, z$ real numbers, and $\mathbf{i}, \mathbf{j}, \mathbf{k}$ pure imaginary numbers that follow the rules $\mathbf{i}^2 = \mathbf{j}^2 = \mathbf{k}^2 = \mathbf{i}\,\mathbf{j}\,\mathbf{k} = -1$ and $\mathbf{i}\,\mathbf{j} = \mathbf{k}$, $\mathbf{j}\,\mathbf{k} = \mathbf{i}$ , $\mathbf{k}\,\mathbf{i} = \mathbf{j}$ (Hamilton, 1847). For unit quaternions, $s^2 + x^2 + y^2 + z^2 = 1$. The composition of two quaternions $\mathbf{q}_1 = s_1 + x_1\mathbf{i} + y_1\,\mathbf{j} + z_1\mathbf{k}$ by $\mathbf{q}_2 = s_2 + x_2\mathbf{i} + y_2\,\mathbf{j} + z_2\,\mathbf{k}$ is obtained by distribution and can be represented on the table

$$\mathbf{q}_3 = \mathbf{q}_1\,\mathbf{q}_2 = \begin{matrix} s_1s_2 & +s_1x_2\mathbf{i} & +s_1y_2\,\mathbf{j} & +s_1z_2\mathbf{k} \\ +x_1s_2\,\mathbf{i} & +x_1x_2\mathbf{i}^2 & +x_1y_2\mathbf{i}\,\mathbf{j} & +x_1z_2\mathbf{i}\,\mathbf{k} \\ +y_1s_2\,\mathbf{j} & +y_1x_2\mathbf{j}\,\mathbf{i} & +y_1y_2\mathbf{j}^2 & +y_1z_2\mathbf{j}\,\mathbf{k} \\ +z_1s_2\,\mathbf{k} & +z_1x_2\mathbf{k}\,\mathbf{i} & +z_1y_2\mathbf{k}\,\mathbf{j} & +z_1z_2\mathbf{k}^2 \end{matrix} = \begin{pmatrix} s_1 \\ x_1 \\ y_1 \\ z_1 \end{pmatrix}^{\mathrm{t}} \begin{pmatrix} 1 & \mathbf{i} & \mathbf{j} & \mathbf{k} \\ \mathbf{i} & -1 & \mathbf{k} & -\mathbf{j} \\ \mathbf{j} & -\mathbf{k} & -1 & \mathbf{i} \\ \mathbf{k} & \mathbf{j} & -\mathbf{i} & -1 \end{pmatrix} \begin{pmatrix} s_2 \\ x_2 \\ y_2 \\ z_2 \end{pmatrix}$$

In other terms, $\mathbf{q}_3 = \mathbf{q}_1^{\mathrm{t}}\,\boldsymbol{Q}\,\mathbf{q}_2$, with $\boldsymbol{Q}$ a tensor that has the form of a 4x4 matrix whose components are made of the scalar 1 and Hamilton's imaginary numbers $\mathbf{i}, \mathbf{j}, \mathbf{k}$ by

$$\boldsymbol{Q} = \begin{pmatrix} 1 & \mathbf{i} & \mathbf{j} & \mathbf{k} \\ \mathbf{i} & -1 & \mathbf{k} & -\mathbf{j} \\ \mathbf{j} & -\mathbf{k} & -1 & \mathbf{i} \\ \mathbf{k} & \mathbf{j} & -\mathbf{i} & -1 \end{pmatrix} \tag{9}$$

Quaternions were initially well received by the scientific community. Maxwell, in his seminal work of 1873, *A Treatise on Electricity and Magnetism*, expressed the electric and magnetic fields as pure vector quaternions, grouping electromagnetic laws into concise quaternion equations. The quaternions were also strongly supported by (Tait, 1890) but they faced a strong opposition from Heaviside and Gibbs, who developed the simpler vector analysis we learn at school. Gibbs showed his aversion to the "*yoke of quaternion*" in a letter to Nature (Gibbs, 1893). For him, the quaternion "*has only occurred from a single man*", and is "*something so separate from all the other branches of learning*" that "*one must give up the progress of [vector algebra]*". To make it very clear he argued that "*in 1886 we have a great deal about imaginaries, and nearly as much as about the quaternion. In 1890 we have nothing about imaginaries, and little about the quaternion*." Maxwell himself was quite ambivalent on the use of quaternions and did not actively participate to the scientific debate. We think that the strong aversion to quaternion mainly relies on the difficulty to understand the imaginary numbers that constitute them. Actually, Hamilton's rules on these imaginaries imply different types of products that are easier to grasp when they are separated. The scalar and cross products were introduced by Gauss and Lagrange, and later formally defined and widely used by Heaviside and Gibbs. They have had a great success in the physics of the 20th century and even now; they gave birth to the concepts of gradient, divergence and curl, and are part of the basics of physics. Hamilton's quaternions made a come-back some decades ago as a specific case of Clifford algebras. Clifford introduced the concept of bi-quaternions (Clifford, 1871), with their algebraic rules that were further generalized into wider mathematical structures. These algebras are nowadays widely used in fundamental physics, general relativity and quantum mechanics. Maxwell equations for example can take the form of a unique equation written in a Clifford algebra (Imaeda, 1995). Complex algebras are powerful because their abstract nature can encompass different algebraic products and many physical properties, and they can be used in a pure logic way without any geometrical representation. They work perfectly well, and are appropriate for "*shut-up and calculate*" approaches, so familiar to quantum mechanics, and quite close to the way AI and brains of brilliant men work. Should it be so? In the case of quaternions, I think not, because we know that unit quaternions already have a geometrical meaning, the rotations, so necessarily, the rules on the imaginaries (symbols) $\mathbf{i}, \mathbf{j}, \mathbf{k}$ on which they are built should also have geometrical meaning. I asked to two AIs (free version) to explain the geometrical meaning of the numbers $\mathbf{i}, \mathbf{j}, \mathbf{k}$ in quaternions, and the answers were not clear nor consistent. Deepseek tells that "$\boldsymbol{i}, \boldsymbol{j}, \boldsymbol{k}$ *are 90°-degree rotation operators in 3D space, around the*

***X**-axis, **Y**-axis, and **Z**-axis"*, respectively. It also tries to give an explanation, by writing "*If you apply* $\boldsymbol{i}$ *twice to a vector (rotate it 90° around **X**, then another 90° around **X**), you flip it completely upside down. That is exactly what multiplying by −1−1 does to a number on a number line. Hence,* $\boldsymbol{i}^2 = -\boldsymbol{i}^2 = -1$. *The same logic applies to **j** and **k***". Other explanations are given based on positive of negative sign of directions "*Think of* 1 *as East,* $-1$ *as West,* $\boldsymbol{i}$ *as North, and* $-\boldsymbol{i}$ *as South*". ChatGPT is vaguer and more cautious than Deepseek: "*You can think of the basis elements like this: Real numbers: one axis. Complex numbers: one real axis + one imaginary axis* $\boldsymbol{i}$. *Quaternions: one real axis + three independent imaginary axes* $\boldsymbol{i}, \boldsymbol{j}, \boldsymbol{k}$. *So* $\boldsymbol{i}, \boldsymbol{j}, \boldsymbol{k}$ *are best understood as three orthogonal imaginary basis elements that encode three-dimensional rotational behavior through their special multiplication rules, rather than as mysterious numbers on their own. Their algebra is precisely what makes quaternions so effective for describing 3D rotations*." All the explanations around the handedness are confusing, such were those attempted by Hamilton in the introduction of his paper (Hamilton, 1847). They did not convince Gibbs or Heaviside and, I must say, they look like raffarinades[1].

The unit complex numbers $\mathbf{c} = a + b\,\mathbf{i}$ with $\sqrt{a^2 + b^2} = 1$ are usually associated with 2D rotations of angle $\alpha$ around the origin with $a = \cos(\alpha)$ and $b = \sin(\alpha)$, which implies that the 2D imaginary number $\mathbf{i}$ is associated with a rotation of $+90°$ (quarter turn), which is apparently in perfect coherence with the fact that $\mathbf{i}^2 = -1$ because twice a rotation of $90°$ is a rotation of angle $180°$, i.e. an inversion in 2D. However, in 3D, in the Cartesian basis $(\mathbf{x}, \mathbf{y}, \mathbf{z})$, the unit quaternion acting on the plane normal to the axis $\mathbf{x}$ is $\mathbf{q} = a + b\,\mathbf{i}$ with $\sqrt{a^2 + b^2} = 1$ and is associated with a 2D rotation of angle $\alpha$ with $a = \cos\left(\frac{\alpha}{2}\right)$ and $b = \sin\left(\frac{\alpha}{2}\right)$. The factor 2 of difference on the angle $\alpha$ between the 2D and 3D imaginary number $\mathbf{i}$ is difficult to explain (and we will see in this work that it should not be so). The pure quaternion number $\mathbf{q} = 0 + 1\,\mathbf{i}$ represents a rotation of angle $180°$ around the $\mathbf{x}$ axis (half turn), so twice this operation is a rotation of $360°$. Writing $\mathbf{i}^2 = -1$ may seem odd because we expect that it would be equal to the identity operation $\mathbf{q} = 1$, but actually the quaternion $\mathbf{q} = \cos\left(\frac{\alpha}{2}\right) + \mathbf{u}$ and the quaternion $-\mathbf{q} = -\cos\left(\frac{\alpha}{2}\right) - \mathbf{u} = \cos\left(\frac{2\pi-\alpha}{2}\right) - (-\mathbf{u})$ represent the same rotation. So, there are actually two units for the quaternions, $1$ and $-1$. We already encountered such ambiguities in the structure of misorientations between variants (Cayron, 2006), and we realized that the explanation was the underlying structure behind the mathematical operations. The product of operators between the variants was shown to be ambiguous because the operators are actually groupoid elements, i.e. arrows formed by pairs of objects similarly "placed" or "orientated". Could it be so for quaternions? Could the structure of quaternions be a groupoid with two units $+1$ *and* $-1$? We will confirm this idea in the rest of the paper by establishing the groupoid composition law associated with quaternions.

[1] Meaningless or self-contradictory sentences. This term is used in France in honour to the former prime minister Jean-Pierre Raffarin who for examples said "*La route est droite, mais la pente est forte*» which can be translated by "*The road is straight but the slope is strong*", and a sentence he said in English with a strong French accent « *The yes needs the no to win against the no*».

This initial purpose of this work was not to find a geometrical meaning of the quaternions and their imaginary numbers. Initially, we just wanted to generalize the Hamiltonian quadratic tensor $\boldsymbol{Q}$ of equations (7)-(8) to non-Cartesian bases by using the product from (1). We realized that the scalar and cross products are intrinsically mixed in this tensor, that is why we willed call it "crossmetric" tensor. It was only when we tried to give a geometrical explanation to its components, that we understood the geometrical meaning of the elementary quaternions (imaginary numbers).

## 2. Cross product in a crystallographic basis written as a symbolic quadratic form

The cross matrix $\boldsymbol{\mathcal{X}}$ (symmetric ) given in equation (7) was introduced in (Cayron, 2026). It permits to calculate the cross product between two vectors $\mathbf{u}$ and $\mathbf{v}$ directly in the crystallographic basis by $\mathbf{u} \times \mathbf{v} = \boldsymbol{\mathcal{X}}\left(\mathbf{u} \underline{\times} \mathbf{v}\right)$ with $\underline{\times}$ the usual Cartesian cross product. This matrix $\boldsymbol{\mathcal{X}}$ makes the calculations easier as it avoids back and forth coordinate changes between the crystallographic basis and arbitrarily defined Cartesian bases. However, the cross product does not have the form of a quadratic product, and thus cannot be combined into a quadratic form similar to that of Hamilton's table (9). The way this issue was solved is detailed below.

As in (Cayron, 2026), the crystal basis $\boldsymbol{\mathcal{B}_c}$ is defined by the three non-colinear crystallographic vectors, $\mathbf{a}, \mathbf{b}, \mathbf{c}$. For two vectors $\mathbf{u}$ and $\mathbf{v}$ in this basis, the cross product $\mathbf{u} \times \mathbf{v}$ is a vector of type $\mathbf{u} \times \mathbf{v} = x\,\mathbf{a} + y\,\mathbf{b} + z\,\mathbf{c}$. Let us use the equation $\mathbf{u} \times \mathbf{v} = \boldsymbol{\mathcal{X}}\left(\mathbf{u} \underline{\times} \mathbf{v}\right)$, and write the column (or row) vectors of the matrix $\boldsymbol{\mathcal{X}}$ by $\boldsymbol{x}_1, \boldsymbol{x}_2, \boldsymbol{x}_3$. Note that since $\boldsymbol{\mathcal{X}} = V\,\boldsymbol{\mathcal{M}}^{-1}$ , these vectors are $\boldsymbol{x}_1 = \mathbf{b} \times \mathbf{c} = V\,\mathbf{a}^*, \boldsymbol{x}_2 = \mathbf{c} \times \mathbf{a} = V\,\mathbf{b}^*, \boldsymbol{x}_3 = \mathbf{a} \times \mathbf{b} = V\,\mathbf{c}^*$. It comes that

$$x = \mathbf{u}^{\mathrm{t}}\,\boldsymbol{\Omega}_1\,\mathbf{v} \qquad (10)$$

$$y = \mathbf{u}^{\mathrm{t}}\,\boldsymbol{\Omega}_2\,\mathbf{v}$$

$$z = \mathbf{u}^{\mathrm{t}}\,\boldsymbol{\Omega}_3\,\mathbf{v}$$

where $\boldsymbol{\Omega}_1$ , $\boldsymbol{\Omega}_2$ and $\boldsymbol{\Omega}_3$ are 3x3 matrices made of the column vectors that are images of $\boldsymbol{x}_1, \boldsymbol{x}_2, \boldsymbol{x}_3$ , respectively, by the three elementary rotation–projection block matrices,

$$\mathbf{L}_1 = \begin{pmatrix} 0 & 0 & 0 \\ 0 & 0 & 1 \\ 0 & -1 & 0 \end{pmatrix}, \mathbf{L}_2 = \begin{pmatrix} 0 & 0 & -1 \\ 0 & 0 & 0 \\ 1 & 0 & 0 \end{pmatrix}, \mathbf{L}_3 = \begin{pmatrix} 0 & 1 & 0 \\ -1 & 0 & 0 \\ 0 & 0 & 0 \end{pmatrix} \qquad (11)$$

such that

$$\boldsymbol{\Omega}_1 = (\mathbf{L}_1\,\boldsymbol{x}_1, \mathbf{L}_2\,\boldsymbol{x}_1, \mathbf{L}_3\,\boldsymbol{x}_1) \qquad (12)$$

$$\boldsymbol{\Omega}_2 = (\mathbf{L}_1\,\boldsymbol{x}_2, \mathbf{L}_2\,\boldsymbol{x}_2, \mathbf{L}_3\,\boldsymbol{x}_2)$$

$$\boldsymbol{\Omega}_3 = (\mathbf{L}_1\,\boldsymbol{x}_3, \mathbf{L}_2\,\boldsymbol{x}_3, \mathbf{L}_3\,\boldsymbol{x}_3)$$

Consequently, the *coordinates* of $\mathbf{u} \times \mathbf{v}$ can be written as a quadratic form, but apparently not the vector $\mathbf{u} \times \mathbf{v}$ itself. However, if we explicitly $\mathbf{u} \times \mathbf{v}$ by its coordinates and its basis vectors $\mathbf{a}, \mathbf{b}, \mathbf{c}$ noted here as symbols

$$\mathbf{u} \times \mathbf{v} = (\mathbf{u}^{\mathrm{t}}\,\boldsymbol{\Omega}_1\,\mathbf{v})\,\mathbf{a} + (\mathbf{u}^{\mathrm{t}}\,\boldsymbol{\Omega}_2\,\mathbf{v})\,\mathbf{b} + (\mathbf{u}^{\mathrm{t}}\,\boldsymbol{\Omega}_3\,\mathbf{v})\,\mathbf{c} \qquad (13)$$

then the cross product vector itself can be written as a quadratic form by

$$\mathbf{u} \times \mathbf{v} = \mathbf{u}^{\mathrm{t}}\,(\boldsymbol{\Omega}_1\,\mathbf{a} + \boldsymbol{\Omega}_2\,\mathbf{b} + \boldsymbol{\Omega}_3\,\mathbf{c})\,\mathbf{v} \tag{14}$$

## 3. The crossmetric tensor and the product of crystallographic quaternions

The scalar part of the product of two crystallographic quaternions is deduced by equation (5) with the metric tensor in a quadratic form $\mathrm{c}_1\mathrm{c}_2 - (\mathbf{u}_1^{\mathrm{t}}\,\boldsymbol{\mathcal{M}}\,\mathbf{u}_2)$. The vectorial part deduced by equation (6) depends on the cross product $\mathrm{c}_2\,\mathbf{u}_1 + \mathrm{c}_1\mathbf{u}_2 + \boldsymbol{\mathcal{X}}\,(\mathbf{u}_1 \underline{\times}\, \mathbf{u}_2)$ and can now be also written as a quadratic form with the symbols $\mathbf{i}_a, \mathbf{i}_b, \mathbf{i}_c$. Both metric and cross tensors can be combined into a unique composition table, that looks like Hamilton's one in equation (9). This table is a 4x4 matrix that will be noted $\boldsymbol{\mathcal{Q}}_{\mathrm{c}}$ (with the subscript c for "crystallographic"); it is

$$\boldsymbol{\mathcal{Q}}_{\mathrm{c}} = \begin{pmatrix} 1 & \mathbf{i}_a & \mathbf{i}_b & \mathbf{i}_c \\ \mathbf{i}_a & \boldsymbol{\mathcal{K}}[1,1] & \boldsymbol{\mathcal{K}}[1,2] & \boldsymbol{\mathcal{K}}[1,3] \\ \mathbf{i}_b & \boldsymbol{\mathcal{K}}[2,1] & \boldsymbol{\mathcal{K}}[2,2] & \boldsymbol{\mathcal{K}}[2,3] \\ \mathbf{i}_c & \boldsymbol{\mathcal{K}}[3,1] & \boldsymbol{\mathcal{K}}[3,2] & \boldsymbol{\mathcal{K}}[3,3] \end{pmatrix} \tag{15}$$

where $\boldsymbol{\mathcal{K}}[i,j]$ are the components in row *i*, column *j* of the 3x3 matrix $\boldsymbol{\mathcal{K}}$ given by

$$\boldsymbol{\mathcal{K}} = -\boldsymbol{\mathcal{M}} + \boldsymbol{\Omega}_1\,\mathbf{i}_a + \boldsymbol{\Omega}_2\,\mathbf{i}_b + \boldsymbol{\Omega}_3\,\mathbf{c} \tag{16}$$

$\boldsymbol{\mathcal{K}}$ is the sum of four 3x3 matrices, each of them associated with one the symbols $\mathbf{i}_a, \mathbf{i}_b, \mathbf{i}_c$, and

The product of two crystallographic quaternions $\mathbf{q}_1 = s_1 + x_1\,\mathbf{i}_a + y_1\,\mathbf{i}_b + z_1\mathbf{c}$ by $\mathbf{q}_2 = s_2 + x_2\,\mathbf{i}_a + y_2\,\mathbf{i}_b + z_2\,\mathbf{i}_c$ is now directly obtained by the quadratic form

$$\mathbf{q}_1\,\mathbf{q}_2 = \begin{pmatrix} s_1 \\ x_1 \\ y_1 \\ z_1 \end{pmatrix}^{\mathrm{t}} \boldsymbol{\mathcal{Q}}_{\mathrm{c}} \begin{pmatrix} s_2 \\ x_2 \\ y_2 \\ z_2 \end{pmatrix} \tag{17}$$

Since the 4x4 matrix $\boldsymbol{\mathcal{Q}}_{\mathrm{c}}$ generalizes the metric tensor for the calculation of the scalar product by a quadratic form, and includes now the cross product, we call it crossmetric tensor. All its components are deduced from the metric tensor $\boldsymbol{\mathcal{M}}$, which reinforces the fundamental importance of $\boldsymbol{\mathcal{M}}$. It should be noted that equation (17) can be used to compose rotational crystallographic quaternions only if these quaternions are normalized, i.e. they should be of type $\mathbf{q} = \mathrm{c} + \mathbf{u}$, with $\mathbf{u} = s\,\tilde{\mathbf{u}}$, with c and s the cosine and sine of the semi-angle of the rotation, $\mathrm{c} = \cos(\frac{\alpha}{2})$ and $\mathrm{s} = \sin(\frac{\alpha}{2})$, and $||\tilde{\mathbf{u}}|| = \sqrt{\tilde{\mathbf{u}}^{\mathrm{t}}\,\boldsymbol{\mathcal{M}}\,\tilde{\mathbf{u}}} = 1$, with $\boldsymbol{\mathcal{M}}$ the metric tensor.

The crossmetric tensor $\boldsymbol{\mathcal{Q}}_{\mathrm{c}}$ can be calculated for any phase once the lattice parameters are known. The metric, cross and crossmetric tensors, $\boldsymbol{\mathcal{M}}, \boldsymbol{\mathcal{X}}, \boldsymbol{\mathcal{Q}}_{\mathrm{c}}$ , respectively, were calculated for the lattices of the six crystal families with Mathematica. They are given in Appendix 1. Note that in the case $a = 1$, we can write $\mathbf{i}_a = \mathbf{i}$, $\mathbf{i}_b = \mathbf{j}$, $\mathbf{i}_c = \mathbf{k}$, and $\boldsymbol{\mathcal{Q}}_{\mathrm{c}} = \boldsymbol{\mathcal{Q}}$ written in equation (9).

The study of the properties of the crossmetric tensor $\boldsymbol{\mathcal{Q}}_{\mathrm{c}}$ would require more work, probably involving more abstract mathematics. We will see that the product of quaternions is actually a groupoid head-tail composition law. In addition, since the concept of Clifford algebra is intimately connected with the theory of quadratic forms and orthogonal transformations, we think that the crossmetric tensor is the explicit (even if symbolic) quadratic matrix associated with the Clifford algebra of crystallographic quaternions. We also observe that the matrices $\mathbf{L}$ in equations (12) are those that define the basis of the Lie algebra SO(3). Many fundamental mathematical notions seem to be intricated in the structure of quaternions. We also note that the matrix $\boldsymbol{\mathcal{Q}}_c$ can be decomposed into a symmetric symbolic matrix and an antisymmetric one by

$$\boldsymbol{\mathcal{Q}}_c = \boldsymbol{\mathcal{Q}}_c^s + \boldsymbol{\mathcal{Q}}_c^a \text{ , with } \boldsymbol{\mathcal{Q}}_c^s = \begin{pmatrix} 1 & \mathbf{i}_a & \mathbf{i}_b & \mathbf{i}_c \\ \mathbf{i}_a & & & \\ \mathbf{i}_b & & -\boldsymbol{\mathcal{M}} & \\ \mathbf{i}_c & & & \end{pmatrix} \text{ and } \boldsymbol{\mathcal{Q}}_c^a = \begin{pmatrix} 0 & 0 & 0 & 0 \\ 0 & 0 & \boldsymbol{\Omega}[1,2] & \boldsymbol{\Omega}[1,3] \\ 0 & \boldsymbol{\Omega}[2,1] & 0 & \boldsymbol{\Omega}[2,3] \\ 0 & \boldsymbol{\Omega}[3,1] & \boldsymbol{\Omega}[3,2] & 0 \end{pmatrix}$$

with $\boldsymbol{\Omega} = \boldsymbol{\Omega}_1\, \mathbf{i}_a + \boldsymbol{\Omega}_2\, \mathbf{i}_b + \boldsymbol{\Omega}_3\, \mathbf{i}_c$.

In crystallography, $\boldsymbol{\mathcal{Q}}_c$ should be such that for any pair of rotational quaternions $\mathbf{q}_1$ and $\mathbf{q}_2$that are symmetries of the crystal, then $\mathbf{q}_1\, \mathbf{q}_2$ should be a quaternion that is also a symmetry of the crystal in order to respect the group structure of the symmetries. These conditions impose strong simplification on the crossmetric tensor $\boldsymbol{\mathcal{Q}}_c$, which explains why the higher the number of symmetries, the simpler $\boldsymbol{\mathcal{Q}}_c$ is, as shown in the tables of Appendix 1. The components obtained for $\boldsymbol{\mathcal{Q}}_c$ for the most general case (triclinic) were already found by (Katrusiak & Le, 2024) by a different method.

## 4. The geometrical meaning of quaternion symbols

### 4.1. The symbols $\mathbf{s}, \mathbf{a}, \mathbf{b}, \mathbf{c}$

Finding the geometrical meaning of the components of $\boldsymbol{\mathcal{Q}}_c$ requires to understand the meaning of Hamilton's symbols $\mathbf{i}, \mathbf{j}, \mathbf{k}$ and their generalization introduced here to non-Cartesian bases, $\mathbf{i}_a, \mathbf{i}_b, \mathbf{i}_c$. They refer to the crystallographic axes $\mathbf{a}, \mathbf{b}, \mathbf{c}$. Crystallographers can refer to the explicit expressions of $\boldsymbol{\mathcal{Q}}_c$ in Appendix 1. According to the triclinic form of the crossmetric tensor, it can be noted that Hamilton's rules generally does not hold for the products between different pure quaternions, for example $\mathbf{i}_a \cdot \mathbf{i}_b$ or $\mathbf{i}_a \cdot \mathbf{i}_c$ . However, Hamilton's square rules for $\mathbf{i}_a, \mathbf{i}_b, \mathbf{i}_c$ are satisfied after normalization, i.e.

$$\frac{\mathbf{i}_a{}^2}{a^2} = \frac{\mathbf{i}_b{}^2}{b^2} = \frac{\mathbf{i}_c{}^2}{c^2} = -1 \tag{18}$$

The demonstration is direct by using $\boldsymbol{\mathcal{Q}}_c$ and the coordinates of the normalized quaternions $\widetilde{\mathbf{i}_a} = (0,\frac{1}{a},0,0)$, $\widetilde{\mathbf{i}_b} = (0,0,\frac{1}{b},0)$ and $\widetilde{\mathbf{i}_c} = (0,0,0,\frac{1}{c})$.

### 4.2. Quaternions as pairs of oriented planes

The key point to understand the imaginary numbers, i.e. the elementary quaternions $\mathbf{i}, \mathbf{j}, \mathbf{k}$, is the fact the rotation based on a quaternion has for angle twice the angle forming the components of the quaternion. This immediately reminded us that reflecting an object twice across two intersecting mirrors results in a rotation by twice the angle between the mirror planes. Here is the bridge between the algebraic quaternions and the geometrical rotations. A quaternion is not really a rotation (an operation); it is a geometrical object than be transformed into an operation, exactly as vectors are objects (arrows) of the 3D space that can be transformed into operations (translations of points) in this space. Geometrically, a quaternion $\mathbf{q}$ of vector $\mathbf{u}$ and angle $\frac{\alpha}{2}$ is the set of all the pairs of oriented planes $(\boldsymbol{m}_1, \boldsymbol{m}_2)$ intersecting along the vector $\mathbf{u} = \boldsymbol{m}_1 \times \boldsymbol{m}_2$ with an angle $\frac{\alpha}{2}$ between them, i.e.

$$\mathbf{q} = (\boldsymbol{m}_1, \boldsymbol{m}_2) \tag{19}$$

The direction of the rotation is always counted positively around $\mathbf{u}$.

The planes $\boldsymbol{m}_1$ and $\boldsymbol{m}_2$ are the mirror planes on which the rotation of angle $\alpha$ around the vector $\mathbf{u}$ is built, as shown in Figure 1a. There is an infinity of pairs of planes $(\boldsymbol{m}_1, \boldsymbol{m}_2)$ defining the same quaternion $\mathbf{q}$, exactly as there is an infinity of pair of points $(\mathbf{A}_1, \mathbf{A}_2)$ defining the same vector $\mathbf{v} = (\mathbf{A}_1, \mathbf{A}_2)$. We used italic for the planes $\boldsymbol{m}_1$ and $\boldsymbol{m}_2$ to remind the fact that these planes can be substituted by other pairs that are images of $(\boldsymbol{m}_1, \boldsymbol{m}_2)$ by any rotation around $\mathbf{u}$ of any angle. Thinking quaternions in terms of pairs of "floating" oriented planes is simple and effective. We already used the fact that quaternions are pairs of mirror planes to demonstrate in the Appendices of (Cayron, 2026) the equations (5)-(6) for the product of rotational crystallographic quaternions. As far as we know, despite the numerous books written on quaternions, despite their use in deep abstract mathematics and in fundamental physics, it seems to be the first time that this simple geometrical interpretation is proposed. It may be hidden inside an existing complex Clifford algebra, but we could not find it. We note that the pair of oriented planes have some similarities with the spinors, but spinors are made of 2x2 Hermitian matrices, with complex entries, whereas here the pair of planes are made of two unit-vectors of the reciprocal space.

As shown in in Figure 1b, the scalar part of a quaternion can be geometrically represented on a trigonometric circle, with the $\mathbf{s}$-axis (scalar axis) horizontal, and the vectorial (imaginary) axis vertical. The point on the circle at the angle $\alpha/2$ marks the angle between the two orientated mirror planes in the pair $(\boldsymbol{m}_1, \boldsymbol{m}_2)$. The quaternion $\mathbf{q} = 1$ represents a rotation of null angle, and the quaternion $\mathbf{q} = -1$ represents a rotation formed by two planes at 180° from each other, and thus a rotation of angle $2 \times 180° = 360°$.

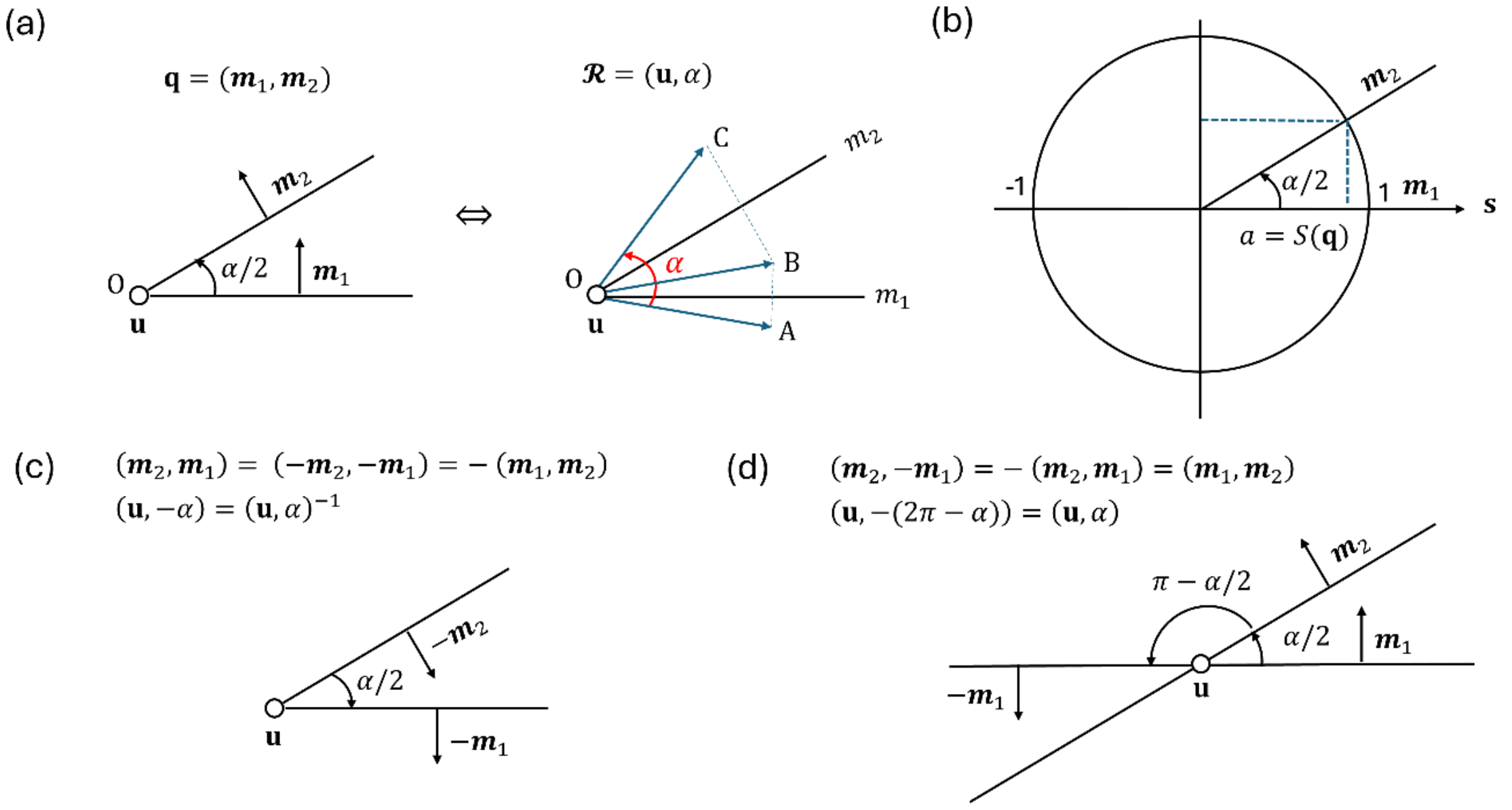

**Figure 1** Geometrical representation of quaternions par pair of oriented planes. (a) Two mirror planes $(\boldsymbol{m}_1, \boldsymbol{m}_2)$ intersecting along the vector $\mathbf{u} = \boldsymbol{m}_1 \times \boldsymbol{m}_2$ (that points toward us) with an angle between them $\alpha/2$ . (b) The scalar part of a quaternion can be represented in a trigonometric circle with $\mathbf{s}$ the scalar axis. Algebraic equalities (c) $(\boldsymbol{m}_2, \boldsymbol{m}_1) = (-\boldsymbol{m}_2, -\boldsymbol{m}_1) = -(\boldsymbol{m}_1, \boldsymbol{m}_2)$ and (d) $(\boldsymbol{m}_2, -\boldsymbol{m}_1) = -(\boldsymbol{m}_2, \boldsymbol{m}_1) = (\boldsymbol{m}_1, \boldsymbol{m}_2)$ resulting from the geometrical definition.

Some basic algebraic composition rules can be derived from the geometrical definition of quaternions. For example, taking the opposite of both planes in the pair does not change the direction of $\mathbf{u}$ because $\mathbf{u} = \boldsymbol{m}_1 \times \boldsymbol{m}_2 = -\boldsymbol{m}_1 \times -\boldsymbol{m}_2$ , and thus does not change the direction of the rotation. In addition, as shown in Figure 1c, starting from the pair $(\boldsymbol{m}_1, \boldsymbol{m}_2)$ and exchanging the planes in the pair and taking their opposite to form $(-\boldsymbol{m}_2, -\boldsymbol{m}_1)$ generates a rotation of opposite sign because $-\boldsymbol{m}_2 \times -\boldsymbol{m}_1 = \boldsymbol{m}_2 \times \boldsymbol{m}_1 = -(\boldsymbol{m}_1 \times \boldsymbol{m}_2)$, thus $(\boldsymbol{m}_2, \boldsymbol{m}_1) = -(\boldsymbol{m}_1, \boldsymbol{m}_2)$. It is also shown in Figure 1d that if $(\boldsymbol{m}_1, \boldsymbol{m}_2)$ is a around $\mathbf{u}$, then $(\boldsymbol{m}_2, -\boldsymbol{m}_1)$ is around $-\mathbf{u}$ because $\boldsymbol{m}_2 \times -\boldsymbol{m}_1 = -(\boldsymbol{m}_1 \times -\boldsymbol{m}_2)$, and its angle of rotation is twice the angle between the planes $(\boldsymbol{m}_2, -\boldsymbol{m}_1)$, $\pi - \frac{\alpha}{2}$, which is thus $-\alpha$. Since turning around $\mathbf{u}$ by an angle $-\alpha$ is equivalent to turning around $-\mathbf{u}$ by an angle $+\alpha$, we can write $(\boldsymbol{m}_2, -\boldsymbol{m}_1) = (\boldsymbol{m}_1, \boldsymbol{m}_2) = -(\boldsymbol{m}_2, \boldsymbol{m}_1)$. In other words, changing simultaneously the direction of rotation and the order of the planes does not affect the rotation.

By applying these equalities for $\boldsymbol{m}_1 = \boldsymbol{m}_2$, it comes that for any plane $\boldsymbol{m}_1$, $(\boldsymbol{m}_1, \boldsymbol{m}_1) = -(\boldsymbol{m}_1, \boldsymbol{m}_1)$. Since this pair represents the rotation of null angle, i.e. the identity quaternion 1 = (1,0,0,0), we can write $(\boldsymbol{m}_1, \boldsymbol{m}_1) = -(\boldsymbol{m}_1, \boldsymbol{m}_1) = 1$. Following the geometrical definition, the quaternion $(\boldsymbol{m}_1, -\boldsymbol{m}_1)$ is the rotation of $2 \times 180° = 360°$; it is thus the quaternion $-1$.

The equalities are summarized by:

$$(\boldsymbol{m}_1, \boldsymbol{m}_2) = (-\boldsymbol{m}_1, -\boldsymbol{m}_2) = (\boldsymbol{m}_2, -\boldsymbol{m}_1) = (-\boldsymbol{m}_2, \boldsymbol{m}_1) = -(\boldsymbol{m}_2, \boldsymbol{m}_1) \qquad (20)$$

In particular:

$$(\boldsymbol{m}_1, \boldsymbol{m}_1) = -(\boldsymbol{m}_1, \boldsymbol{m}_1) = 1 \qquad (21)$$

$$(\boldsymbol{m}_1, -\boldsymbol{m}_1) = (-\boldsymbol{m}_1, \boldsymbol{m}_1) = -1$$

### 4.3. Head-tail composition of quaternions

Assimilating a quaternion to an infinite set of pairs of oriented planes (that are the mirror planes on which the rotation is built) allows us to compose quaternions geometrically by using the elementary head-tail rule, also called source-target groupoid rule. Two quaternions $\mathbf{q} = (\boldsymbol{m}_1, \boldsymbol{m}_2)$ and $\mathbf{r} = (\boldsymbol{m}_3, \boldsymbol{m}_4)$ of axis $\mathbf{u}$ and $\mathbf{v}$, respectively, can be geometrically composed when the floating head plane of $\mathbf{q}$ is in coincidence with the floating tail plane of $\mathbf{r}$, i.e. $\boldsymbol{m}_2 = \boldsymbol{m}_3$, and the resulting quaternion is $\mathbf{q}\,\mathbf{r} = (\boldsymbol{m}_1, \boldsymbol{m}_4)$. The plane $\boldsymbol{m}_2 = \boldsymbol{m}_3$ is the plane spanned by $\mathbf{u}$ and $\mathbf{v}$, i.e. $\boldsymbol{m}_2 = \boldsymbol{m}_3 = \mathbf{u} \times \mathbf{v}$, or any plane containing $\mathbf{u}$ if $\mathbf{u}$ and $\mathbf{v}$ are parallel. The head-tail (or source-target) law is often encountered in mathematics. Vectors and coordinate transformation matrices are composed in this way. The vector $\mathbf{u} = (\mathbf{A}_1, \mathbf{A}_2)$ can be geometrically composed with the vector $\mathbf{v} = (\mathbf{A}_3, \mathbf{A}_4)$ by placing the floating head point of $\mathbf{u}$ in coincidence with the floating tail point of $\mathbf{v}$, i.e. $\mathbf{A}_3 = \mathbf{A}_4$, and the result is the vector $\mathbf{u}+\mathbf{v}=(\mathbf{A}_1,\mathbf{A}_2)+(\mathbf{A}_2,\mathbf{A}_4)=(\mathbf{A}_1,\mathbf{A}_4)$. The coordinate transformation matrix $\mathbf{P}$ going from a floating basis $\boldsymbol{\mathcal{B}}_1$ to a floating basis $\boldsymbol{\mathcal{B}}_2$, $\mathbf{Q}=[\boldsymbol{\mathcal{B}}_1\to\boldsymbol{\mathcal{B}}_2]$ can be composed with $\mathbf{Q}=[\boldsymbol{\mathcal{B}}_3\to\boldsymbol{\mathcal{B}}_4]$ if $\boldsymbol{\mathcal{B}}_2=\boldsymbol{\mathcal{B}}_3$, and the result is the matrix $\mathbf{P}\,\mathbf{Q}=[\boldsymbol{\mathcal{B}}_1\to\boldsymbol{\mathcal{B}}_2][\boldsymbol{\mathcal{B}}_2\to\boldsymbol{\mathcal{B}}_4]=[\boldsymbol{\mathcal{B}}_1\to\boldsymbol{\mathcal{B}}_4]$. We already encountered this law to define the misorientations between the variants as arrows (double-cosets) between variants (simple-cosets) forming a groupoid structure (Cayron, 2006).

In summary, the composition of quaternions follows the groupoid head-tail rule. This rule can be naturally written by indexing the quaternions as $\mathbf{q}_{12}=(\boldsymbol{m}_1,\boldsymbol{m}_2)$, $\mathbf{q}_{23}=(\boldsymbol{m}_2,\boldsymbol{m}_3)$, such that their product is

$$(\boldsymbol{m}_1,\boldsymbol{m}_2)\,(\boldsymbol{m}_2,\boldsymbol{m}_3)=(\boldsymbol{m}_1,\boldsymbol{m}_3) \quad (22)$$

or equivalently $\mathbf{q}_{12}\,\mathbf{q}_{23}=\mathbf{q}_{13}$

### 4.4. The meaning of the elementary Cartesian quaternions $\mathbf{i},\mathbf{j},\mathbf{k}$

Now, we can come back to Hamilton's rules $\mathbf{i}^2=\mathbf{j}^2=\mathbf{k}^2=-1$, $\mathbf{i}\,\mathbf{j}=\mathbf{k}$, $\mathbf{j}\,\mathbf{k}=\mathbf{i}$, $\mathbf{k}\,\mathbf{i}=\mathbf{j}$, and $\mathbf{i}\,\mathbf{j}\,\mathbf{k}=-1$. Let us consider the elementary quaternions $\mathbf{i},\mathbf{j},\mathbf{k}$, or imaginary numbers, as pairs of floating oriented planes. Since these quaternions have a null scalar part and that the scalar is $\cos(\frac{\alpha}{2})$, the symbols $\mathbf{i},\mathbf{j},\mathbf{k}$ represent half-turn rotations around the axis $\mathbf{x},\mathbf{y},\mathbf{z}$, respectively. If we consider the Cartesian basis $(\mathbf{x},\mathbf{y},\mathbf{z})$ and the orientated planes $\boldsymbol{m}_x$, $\boldsymbol{m}_y$ and $\boldsymbol{m}_z$ as shown in Figure 2, we can write the quaternions by a representative in the set of equivalent pairs of planes, for example:

$$\begin{aligned}\mathbf{i}&=(\boldsymbol{m}_y,-\boldsymbol{m}_z)\\ \mathbf{j}&=(\boldsymbol{m}_z,-\boldsymbol{m}_x)\\ \mathbf{k}&=(\boldsymbol{m}_x,-\boldsymbol{m}_y)\end{aligned} \quad (23)$$

Note that to be valid, the definition (23) imposes to take the clockwise direction as the positive direction of rotation. This results from the passive meaning we have attributed to the quaternions, with a left-to-right product. Using an active meaning with a right-to-left product would lead to impose a counterclockwise direction of rotation.

Hamilton probably thought that the rules he discovered for the quaternions were fundamental in the way that they do not result from more elementary ones, but we show here that this is not the case. Actually, Hamilton's rules result from the head-tail composition rule (22); as illustrated in Figure 2a. In particular:

a) $\mathbf{i}^2=(\boldsymbol{m}_y,-\boldsymbol{m}_z)(\boldsymbol{m}_y,-\boldsymbol{m}_z)=-(\boldsymbol{m}_y,\boldsymbol{m}_z)(\boldsymbol{m}_z,\boldsymbol{m}_y)=-(\boldsymbol{m}_y,\boldsymbol{m}_y)=-1$

b) $\mathbf{i}\,\mathbf{j}=(\boldsymbol{m}_y,-\boldsymbol{m}_z)(\boldsymbol{m}_z,-\boldsymbol{m}_x)=(\boldsymbol{m}_y,\boldsymbol{m}_x)=(\boldsymbol{m}_x,-\boldsymbol{m}_y)=\mathbf{k}$

c) $\mathbf{j}\,\mathbf{i}=(\boldsymbol{m}_z,-\boldsymbol{m}_x)(\boldsymbol{m}_y,-\boldsymbol{m}_z)=(-\boldsymbol{m}_x,\boldsymbol{m}_z)(\boldsymbol{m}_z,-\boldsymbol{m}_y)=(-\boldsymbol{m}_x,-\boldsymbol{m}_y)=(\boldsymbol{m}_x,\boldsymbol{m}_y)=-\mathbf{k}$

and $\mathbf{j}^2=\mathbf{k}^2=-1$, $\mathbf{j}\,\mathbf{k}=\mathbf{i}$, $\mathbf{k}\,\mathbf{i}=\mathbf{j}$ follow by permutation, and $\mathbf{i}\,\mathbf{j}\,\mathbf{k}=-1$ by combination.

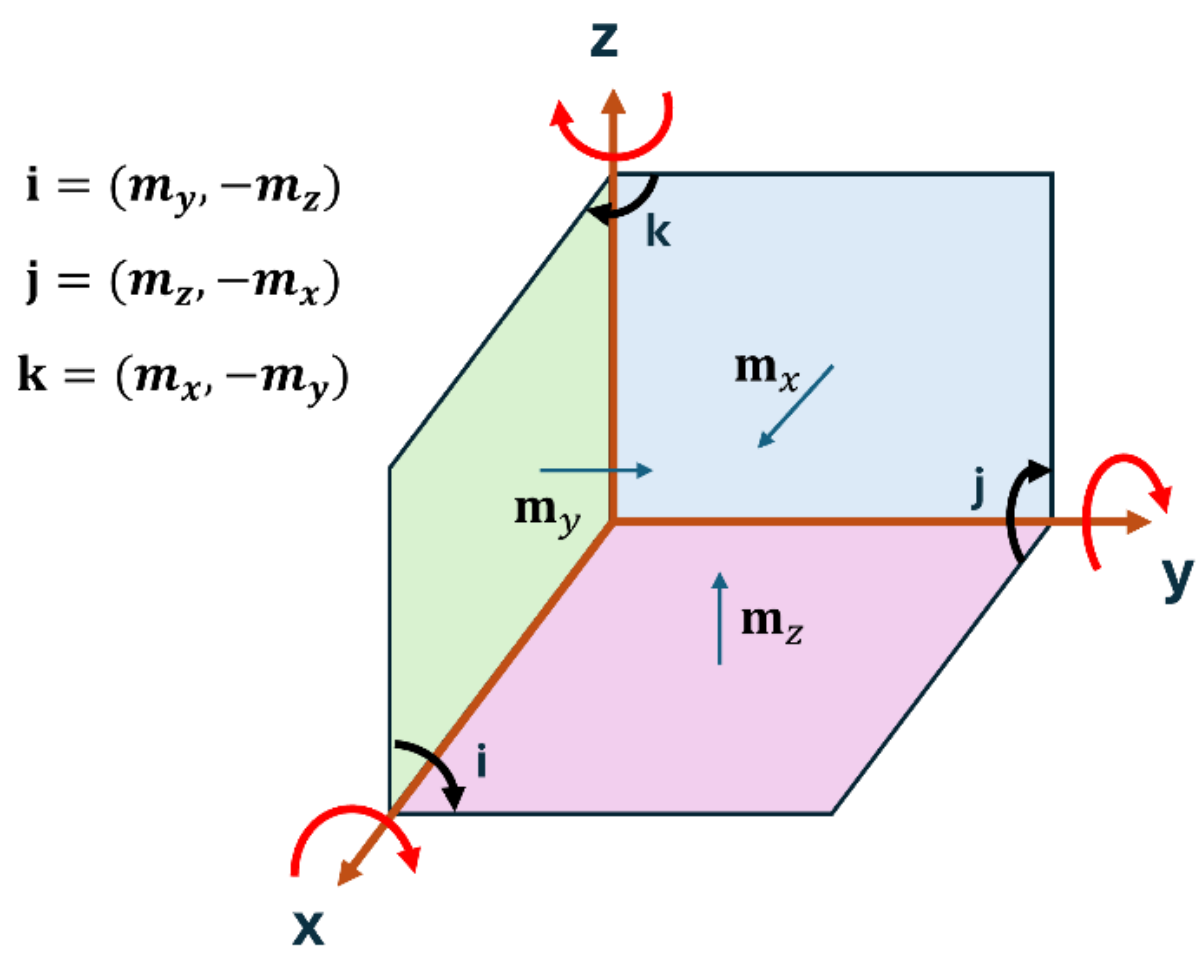

**Figure 2** Geometrical meaning of elementary quaternions, (a) $\mathbf{i}, \mathbf{j}, \mathbf{k}$ for a Cartesian basis. The direction of rotation is clockwise.

### 4.5. The meaning of the 2D imaginary number $\mathbf{i}$

In order to rationalize the meaning of the imaginary numbers $\mathbf{i}, \mathbf{j}, \mathbf{k}$ used in quaternions with the number $\mathbf{i}$ used for complex numbers, we think that the 2D imaginary number $\mathbf{i}$ solution of the equation $\mathbf{x}^2 = -1$ should not be imagined as a rotation of 90°, as it is usually believed, but as the set of pairs of floating perpendicular and oriented planes (actually lines because of the 2D space), which could be written $\mathbf{i} = (\boldsymbol{m}_x, -\boldsymbol{m}_y)$ for a Cartesian basis $(\mathbf{x}, \mathbf{y})$. As for quaternion, the rotation associated with $\mathbf{i}$ is a rotation of twice +90°, i.e. 180°. As for quaternion, $\mathbf{i}^2 = (\boldsymbol{m}_x, -\boldsymbol{m}_y)(\boldsymbol{m}_x, -\boldsymbol{m}_y) = (\boldsymbol{m}_x, -\boldsymbol{m}_y)(-\boldsymbol{m}_y, -\boldsymbol{m}_x) = (\boldsymbol{m}_x, -\boldsymbol{m}_x) = -1$. Since in 2D, all the rotations turn around the origin, the isomorphism between the rotations of angle $\alpha/2$ and those of angle $\alpha$ has masked the deep nature of the number $\mathbf{i}$. The complex number $\cos\alpha + \sin\alpha\, \mathbf{i}$ does *not* represent a rotation of angle α; it represents two planes forming an angle α between them, and thus a rotation of angle 2α. All our old geometrical representations of complex numbers proposed by Jean-Pierre Argand and Sophus Lie in 1897 are still valid, but we have to realize that what we represent in the trigonometric circle to show a unit complex number is the target mirror plane, exactly as shown in Figure 1b for quaternions.

### 4.6. The meaning of the elementary crystallographic quaternions

Initially we thought that the elementary crystallographic quaternions could be defined as pairs of planes $\boldsymbol{m}_a = (\mathbf{b}, \mathbf{c})$, $\boldsymbol{m}_b = (\mathbf{c}, \mathbf{b})$, $\boldsymbol{m}_c = (\mathbf{a}, \mathbf{b})$, but we realized that this definition would disagree with the composition rule (18). We thus came to conclude that the unit crystallographic quaternion $\tilde{\mathbf{i}}_a$ is actually the set of all pairs of floating planes $(\boldsymbol{m}_a^-, \boldsymbol{m}_a^+)$ such that the angle between $\boldsymbol{m}_a^-$ and $\boldsymbol{m}_a^+$ is +90° (positive quarter turn), and so for the unit quaternions $\tilde{\mathbf{i}}_b$, $\tilde{\mathbf{i}}_c$. As for the other quaternions, the planes in the pairs are "floating planes", and any pair of planes that are images of $(\boldsymbol{m}_a^-, \boldsymbol{m}_a^+)$ by a rotation of positive angle around $\tilde{\mathbf{a}}$ would also form the same quaternion $\tilde{\mathbf{i}}_a$. We thus note

$$\tilde{\mathbf{i}}_a = (\boldsymbol{m}_a^-, \boldsymbol{m}_a^+) \qquad (24)$$
$$\tilde{\mathbf{i}}_b = (\boldsymbol{m}_b^-, \boldsymbol{m}_b^+)$$

$$\tilde{\mathbf{i}}_c = (\boldsymbol{m}_c^-, \boldsymbol{m}_c^+)$$

It is direct to see that the rule (18), ${\tilde{\mathbf{i}}_a}^2 = {\tilde{\mathbf{i}}_b}^2 = {\tilde{\mathbf{i}}_c}^2 = -1$, already found by using the crossmetric tensor also results from the geometrical rules (20). The elementary crystallographic quaternions are shown in Figure 3a.

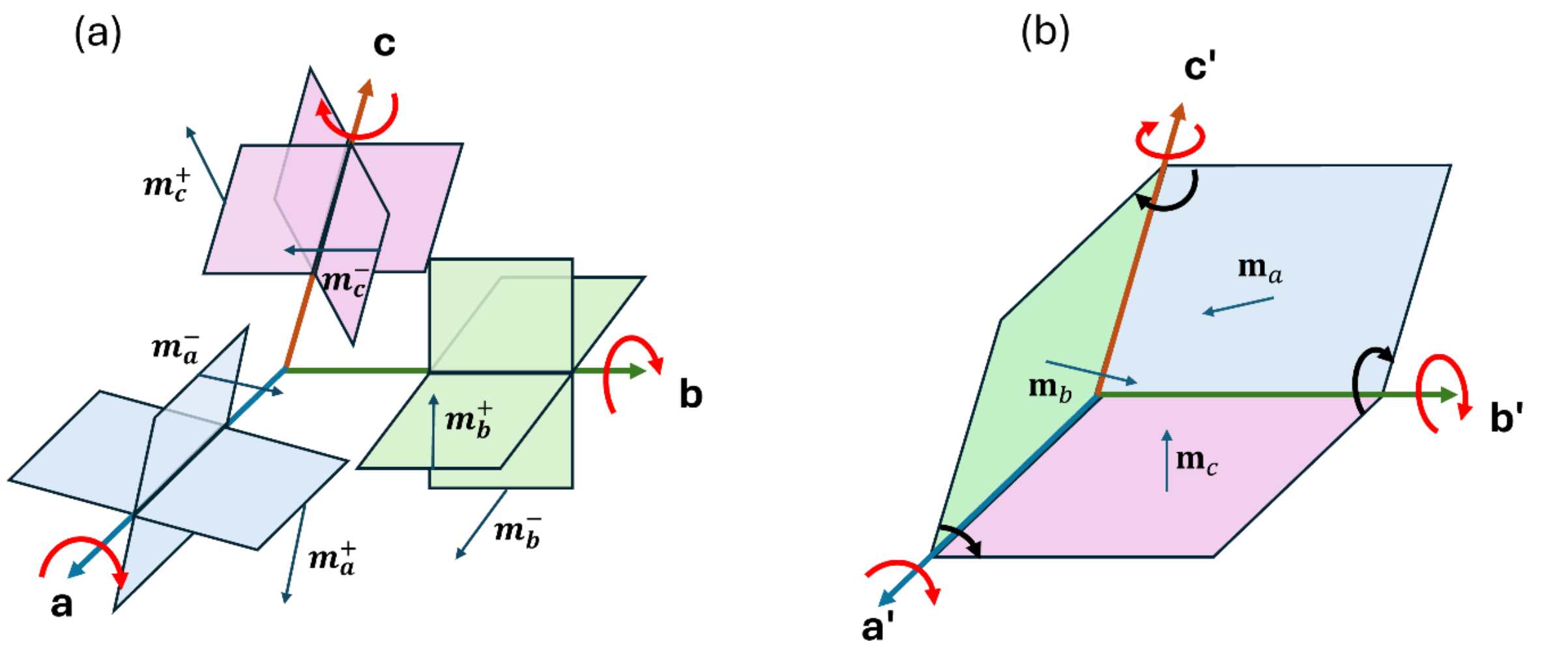

**Figure 3** Geometrical representation of the elementary and complementary crystallographic quaternions as pairs of oriented planes. (a) Elementary quaternions $\tilde{\mathbf{i}}_a, \tilde{\mathbf{i}}_b, \tilde{\mathbf{i}}_c$ formed by pairs of perpendicular and oriented planes intersecting along the axes **a**, **b**, **c**, respectively. (b) Complementary quaternions $\tilde{\mathbf{i}}_a', \tilde{\mathbf{i}}_b', \tilde{\mathbf{i}}_c'$ formed by pairs of crystallographic planes $\boldsymbol{m_a} = (\mathbf{b}, \mathbf{c})$, $\boldsymbol{m_b} = (\mathbf{c}, \mathbf{b})$, $\boldsymbol{m_c} = (\mathbf{a}, \mathbf{b})$. The direction of rotation is clockwise.

Now, we wonder: what are the quaternions formed on the pairs of planes $\boldsymbol{m_a} = (\mathbf{b}, \mathbf{c})$, $\boldsymbol{m_b} = (\mathbf{c}, \mathbf{a})$, $\boldsymbol{m_c} = (\mathbf{a}, \mathbf{b})$? The Miller indices of these planes are those of the reciprocal vectors $\mathbf{a}^* = (1,0,0)$, $\mathbf{b}^*, = (0,1,0)$, $\mathbf{c}^* = (0,01)$. We note them

$$\tilde{\mathbf{i}}_a' = (\boldsymbol{m}_b, \boldsymbol{m}_c) \quad (25)$$
$$\tilde{\mathbf{i}}_b' = (\boldsymbol{m}_c, \boldsymbol{m}_a)$$
$$\tilde{\mathbf{i}}_c' = (\boldsymbol{m}_a, \boldsymbol{m}_b)$$

They are shown in Figure 3b.We call them “complementary” quaternions, because they complete Hamilton’s rules; indeed, their definition allows us to establish without calculations that

$$\tilde{\mathbf{i}}_a'\tilde{\mathbf{i}}_b' = \tilde{\mathbf{i}}_c', \quad \tilde{\mathbf{i}}_b'\tilde{\mathbf{i}}_c' = \tilde{\mathbf{i}}_a', \quad (26)$$
$$\tilde{\mathbf{i}}_c'\tilde{\mathbf{i}}_a' = \tilde{\mathbf{i}}_b', \quad \tilde{\mathbf{i}}_b'\tilde{\mathbf{i}}_a' = -\tilde{\mathbf{i}}_c', \quad \tilde{\mathbf{i}}_c'\tilde{\mathbf{i}}_b' = -\tilde{\mathbf{i}}_a'$$
$$\tilde{\mathbf{i}}_a'\tilde{\mathbf{i}}_c' = -\tilde{\mathbf{i}}_b', \quad \tilde{\mathbf{i}}_a'\tilde{\mathbf{i}}_b'\tilde{\mathbf{i}}_c' = -1$$

The crystallographic quaternions $\tilde{\mathbf{i}}_a, \tilde{\mathbf{i}}_b, \tilde{\mathbf{i}}_c$ and their complementary $\tilde{\mathbf{i}}_a', \tilde{\mathbf{i}}_b', \tilde{\mathbf{i}}_c'$ allow establishing Hamilton-like rules of composition with the equations (18) and (26). The six quaternions could thus be used to build an algebra, but this subject is out of the scope of the paper, and beyond the capability of the author.

Cartesian quaternions constitute a specific case of crystallographic quaternion for which the complementary quaternions are the same as the elementary quaternions, i.e. $\mathbf{i} = \mathbf{i}', \mathbf{j} = \mathbf{j}', \mathbf{k} = \mathbf{k}'$.

### 4.7. Geometrical meaning of quaternions and components of the crossmetric tensor

The geometrical representation of quaternions permits to understand the components of the 4x4 quadratic matrix $\boldsymbol{\mathcal{Q}}_c$. For example, according to Appendix 1, in an hexagonal lattice, the product of the two crystallographic quaternions $\mathbf{i}_a\,\mathbf{i}_c = \begin{pmatrix}0\\1\\0\\0\end{pmatrix}^t \boldsymbol{\mathcal{Q}}_c \begin{pmatrix}0\\0\\1\\0\end{pmatrix} = \boldsymbol{\mathcal{Q}}_c[2,4] = \boldsymbol{\mathcal{K}}[1,3]$ is $\boldsymbol{\mathcal{Q}}_c[2,4] = -\frac{c\,(\mathbf{i}_a+2\,\mathbf{i}_b)}{\sqrt{3}}$. This quaternion can also be written $-\frac{ac\,(\tilde{\mathbf{i}}_a+2\,\tilde{\mathbf{i}}_b)}{\sqrt{3}}$. It is this a rotation of 180° around the axis $-\frac{(\mathbf{a}+2\,\mathbf{b})}{\sqrt{3}\,a}$. This result is geometrically illustrated in Figure 4a. It is shown that in order to compose the quaternion $\mathbf{i}_a$ with the quaternion $\mathbf{i}_c$, the target plane of $\mathbf{i}_a$ is positioned in coincidence with the source plane of the quaternion $\mathbf{i}_c$ , and we write $\mathbf{i}_a = a\,(\boldsymbol{m}_a^-, \boldsymbol{m}_a^+)$ and $\mathbf{i}_c = c\,(\boldsymbol{m}_c^-, \boldsymbol{m}_c^+)$ with $\boldsymbol{m}_c^- = -\boldsymbol{m}_a^+$, thus the resulting quaternion is $\mathbf{i}_a\,\mathbf{i}_c = a\,c\,(\boldsymbol{m}_a^-, -\boldsymbol{m}_c^+) = -a\,c\,(\boldsymbol{m}_a^-, \boldsymbol{m}_c^+)$. Its vectorial part is the intersection of the planes $\boldsymbol{m}_a^-$ and $\boldsymbol{m}_c^+$ , which is simply $\mathbf{a} + 2\,\mathbf{b}$. After normalization by the metric tensor, this vector is $-\frac{(\mathbf{a}+2\,\mathbf{b})}{\sqrt{3}\,a}$, as expected. The scalar part of $\mathbf{i}_a\,\mathbf{i}_c$ is null because $\boldsymbol{m}_a^-$ and $\boldsymbol{m}_c^+$ are perpendicular.

Similarly, for a monoclinic lattice, according to Appendix 1, the component $\mathbf{i}_a\,\mathbf{i}_c = \boldsymbol{\mathcal{Q}}_c[2,4] = -a\,c\,\left(\frac{\sin\beta}{b}\,\mathbf{i}_b + \cos\beta\right)$. This result is illustrated geometrically in Figure 4b. Here again, we write $\mathbf{i}_a = a\,(\boldsymbol{m}_a^-, \boldsymbol{m}_a^+)$ and $\mathbf{i}_c = c\,(\boldsymbol{m}_c^-, \boldsymbol{m}_c^+)$ with $\boldsymbol{m}_c^+ = -\boldsymbol{m}_a^-$, thus the resulting quaternion is $\mathbf{i}_a\,\mathbf{i}_c = a\,c\,(\boldsymbol{m}_a^+, \boldsymbol{m}_c^-) = -a\,c\,(\boldsymbol{m}_c^-, \boldsymbol{m}_a^+)$ where the vectorial part resulting from the intersection of the planes $\boldsymbol{m}_c^-$ and $\boldsymbol{m}_a^+$ is the vector $\mathbf{b}$. The angle of the quaternion, i.e. the semiangle of rotation, is the angle between $\boldsymbol{m}_c^-$ and $\boldsymbol{m}_a^+$, i.e. the monoclinic angle $\beta$. Consequently, the scalar part of the resulting quaternion $\mathbf{i}_a\,\mathbf{i}_c$ is $\cos\beta$ and its vectorial part is $\frac{\sin\beta}{b}\,\mathbf{b}$. The result $\mathbf{i}_a\,\mathbf{i}_c = a\,c\,\left(\frac{\sin\beta}{b}\,\mathbf{i}_b + \cos\beta\right)$ is thus "proved" geometrically.

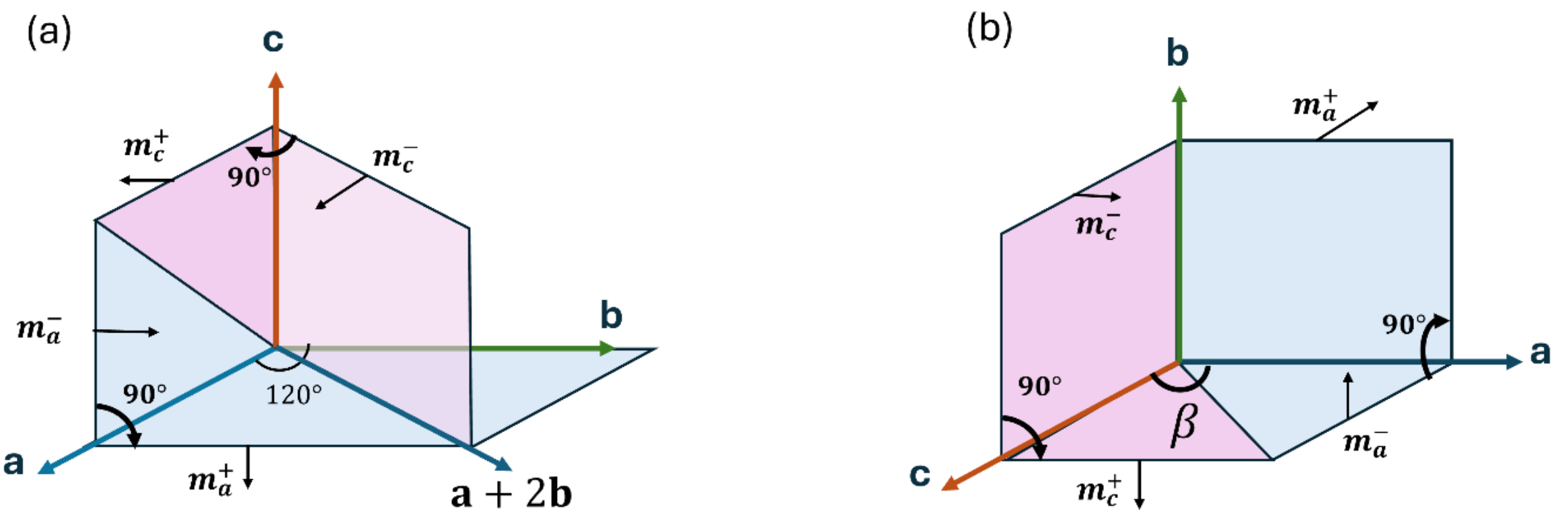

**Figure 4** Geometrical representation of the diagonal components of the crossmetric tensor $\boldsymbol{\mathcal{Q}}_c$. (a) Component$\mathbf{i}_a\,\mathbf{i}_c = \boldsymbol{\mathcal{Q}}_c[2,4]$ for a hexagonal structure. (b) Component $\mathbf{i}_a\,\mathbf{i}_c = \boldsymbol{\mathcal{Q}}_c[2,4]$ for a monoclinic structure of angle $\beta$. The direction of rotation is clockwise.

## 5. Conclusions

We showed in a previous paper (Cayron, 2026) that Cartesian quaternions can be broaden to non-Cartesian bases by generalizing the scalar and cross product with the help of the metric and cross tensors. We introduced the cross tensor to calculate cross products directly in the crystallographic basis, similarly as for the metric tensor for the scalar products. For Cartesian bases, the elementary quaternions are $\mathbf{i}, \mathbf{j}, \mathbf{k}$, and Hamilton's rules on them are $\mathbf{i}^2 = \mathbf{j}^2 = \mathbf{k}^2 = \mathbf{i}\,\mathbf{j}\,\mathbf{k} = -1$ and $\mathbf{i}\,\mathbf{j} = \mathbf{k}$, $\mathbf{j}\,\mathbf{k} = \mathbf{i}$ , $\mathbf{k}\,\mathbf{i} = \mathbf{j}$ . These rules can be condensed into a unique 4x4 matrix $\boldsymbol{Q}$ made of symbols $\mathbf{i}, \mathbf{j}, \mathbf{k}$ that makes the product of two Cartesian quaternions a quadratic form. We showed in the present study that an equivalent matrix $\boldsymbol{Q}_{\mathrm{c}}$ exists for the crystallographic quaternions, and we called it "crossmetric tensor" because it combines both the metric and the cross tensors. It can be calculated for any lattice uniquely from the metric tensor. The symbols used to write it are the crystallographic imaginary numbers $\mathbf{i}_a, \mathbf{i}_b, \mathbf{i}_c$; they are also, after normalization, the elementary crystallographic quaternions that generalize the usual$\mathbf{i}, \mathbf{j}, \mathbf{k}$ elementary Cartesian quaternions. The 4x4 matrices $\boldsymbol{Q}_{\mathrm{c}}$ were explicitly given for the six crystal families.

This work allowed us to understand that a unit quaternion of scalar part $\cos(\frac{\alpha}{2})$ and vector part $\sin\left(\frac{\alpha}{2}\right)\,\tilde{\mathbf{u}}$ can be geometrically represented by an infinity of pairs of oriented planes $(\boldsymbol{m}_1, \boldsymbol{m}_2)$ intersecting along $\mathbf{u}$ with an angle $\frac{\alpha}{2}$ between them. These planes are mirror planes from which the rotation of axis $\mathbf{u}$ and angle $\alpha$ are deduced. Since the planes associated with a quaternion are not unique, we called them "floating" planes. The floating plane at the left in the pair is called source (or tail) plane, and that at the right target (or head) plane. The composition of two quaternions is geometrically obtained by placing in coincidence the target plane of the first quaternion with the source plane of the second quaternion, according to the source-target groupoid composition law $(\boldsymbol{m}_1, \boldsymbol{m}_2)\,(\boldsymbol{m}_2, \boldsymbol{m}_3) = (\boldsymbol{m}_1, \boldsymbol{m}_3)$. By considering the components of the crossmetric tensor, we understood that the elementary quaternions $\tilde{\mathbf{i}}_a, \tilde{\mathbf{i}}_b, \tilde{\mathbf{i}}_c$ are geometrically represented by the infinite set of pairs of perpendicular planes intersecting along the axis $\mathbf{a}, \mathbf{b}$, or $\mathbf{c}$, respectively. They obey only the square Hamilton's rule in the way that $\tilde{\mathbf{i}}_a^{\,2} = \tilde{\mathbf{i}}_b^{\,2} = \tilde{\mathbf{i}}_c^{\,2} = -1$. Three "complementary" quaternions $\tilde{\mathbf{i}}_a', \tilde{\mathbf{i}}_b', \tilde{\mathbf{i}}_c'$ were also introduced. These quaternions are based on the pairs of crystallographic planes $\boldsymbol{m}_{\boldsymbol{a}}$, $\boldsymbol{m}_{\boldsymbol{b}}$, $\boldsymbol{m}_{\boldsymbol{c}}$. Their composition rules result immediately from the definition; they are Hamilton's rules on the bi or tri-composition of different quaternions, $\tilde{\mathbf{i}}_a'\tilde{\mathbf{i}}_b' = \tilde{\mathbf{i}}_c'$, $\tilde{\mathbf{i}}_b'\tilde{\mathbf{i}}_c' = \tilde{\mathbf{i}}_a'$, $\tilde{\mathbf{i}}_c'\tilde{\mathbf{i}}_a' = \tilde{\mathbf{i}}_b'$, $\tilde{\mathbf{i}}_b'\tilde{\mathbf{i}}_a' = -\tilde{\mathbf{i}}_c'$, $\tilde{\mathbf{i}}_c'\,\tilde{\mathbf{i}}_b' = -\tilde{\mathbf{i}}_a'$, $\tilde{\mathbf{i}}_a'\,\tilde{\mathbf{i}}_c' = -\tilde{\mathbf{i}}_b'$, $\tilde{\mathbf{i}}_a'\,\tilde{\mathbf{i}}_b'\,\tilde{\mathbf{i}}_c' = -1$. It was also shown that the head-tail rule allows to determine the components of the crossmetric tensor by geometry.

The elementary Cartesian quaternions $\mathbf{i}, \mathbf{j}, \mathbf{k}$ appear as a specific case of crystallographic quaternions. They are the pairs formed on the perpendicular faces of the cube $\boldsymbol{m}_{\boldsymbol{x}}$, $\boldsymbol{m}_{\boldsymbol{y}}$, $\boldsymbol{m}_{\boldsymbol{z}}$ such that $\mathbf{i} = (\boldsymbol{m}_y, -\boldsymbol{m}_z)$, $\mathbf{j} = (\boldsymbol{m}_z, -\boldsymbol{m}_x)$, $\mathbf{k} = (\boldsymbol{m}_x, -\boldsymbol{m}_y)$. Since these planes intersect along the $\boldsymbol{x}$, y and z axis, respectively, the elementary and complementary quaternions are equal, i.e. $\mathbf{i} = \mathbf{i}', \mathbf{j} = \mathbf{j}', \mathbf{k} = \mathbf{k}'$, which explains why the square, bi and tri-product Hamilton's rules are all satisfied with only three quaternions $\mathbf{i}, \mathbf{j}, \mathbf{k}$ and not six. In order to rationalize the 2D imaginary number $\mathbf{i}$ and the 3D imaginary numbers $\mathbf{i}, \mathbf{j}, \mathbf{k}$, one should forget the idea that the 2D imaginary number $\mathbf{i}$ would be rotation of +90° around the origin, and consider it as the pairs of perpendicular planes $(\boldsymbol{m}_x, -\boldsymbol{m}_y)$ representing a rotation of 180°.

**Acknowledgements** I acknowledge Prof. Roland Logé, chief of LMTM, the lab where I work as a microscopist, metallurgist, and crystallographer; and PX group, the company that chaired the lab for ten years.

**Note** AI was used only to get a better knowledge on the common view of quaternions, and to get an idea of the main concepts on which Clifford algebras are built. We did not use AI for any other purpose. The figures were drawn with Power Point, which can explain their probable inaccuracy. The symbolic calculations were made with Mathematica; the coordinates along the axes $\mathbf{i}_a$ $\mathbf{i}_c$ were identified with the function "Coefficient". The Mathematica program allows calculating the product of crystallographic quaternions by using the generalized scalar or cross products given in equations (5) and (6), and by calculating the crossmetric tensor by equation (15) to form the quadratic product by equation (17). This program is available upon request.

**Appendix 1.** **Metric, cross and crossmetric matrices, $\mathcal{M}$, $\mathcal{X}$, $\mathcal{Q}_c$ , respectively, for the six crystal families**

| Structure with lattice parameters | Metric $\boldsymbol{\mathcal{M}}$ | Cross $\boldsymbol{\mathcal{X}}$ | Crossmetric $\boldsymbol{\mathcal{Q}}_c$ |
|---|---|---|---|
| **Cubic**, $a$ | $\begin{pmatrix} a^2 & 0 & 0 \\ 0 & a^2 & 0 \\ 0 & 0 & a^2 \end{pmatrix}$ | $\begin{pmatrix} a & 0 & 0 \\ 0 & a & 0 \\ 0 & 0 & a \end{pmatrix}$ | $\begin{pmatrix} 1 & \mathbf{i}_a & \mathbf{i}_b & \mathbf{i}_c \\ \mathbf{i}_a & -a^2 & a\,\mathbf{i}_c & -a\,\mathbf{i}_b \\ \mathbf{i}_b & -a\,\mathbf{i}_c & -a^2 & a\,\mathbf{i}_a \\ \mathbf{i}_c & a\,\mathbf{i}_b & -a\,\mathbf{i}_a & -a^2 \end{pmatrix}$ |
| **Tetragonal**, $a, c$ | $\begin{pmatrix} a^2 & 0 & 0 \\ 0 & a^2 & 0 \\ 0 & 0 & c^2 \end{pmatrix}$ | $\begin{pmatrix} c & 0 & 0 \\ 0 & c & 0 \\ 0 & 0 & \frac{a^2}{c} \end{pmatrix}$ | $\begin{pmatrix} 1 & \mathbf{i}_a & \mathbf{i}_b & \mathbf{i}_c \\ \mathbf{i}_a & -a^2 & \frac{a^2}{c}\mathbf{i}_c & -c\,\mathbf{i}_b \\ \mathbf{i}_b & -\frac{a^2}{c}\mathbf{i}_c & -a^2 & c\,\mathbf{i}_a \\ \mathbf{i}_c & c\,\mathbf{i}_b & -c\,\mathbf{i}_a & -c^2 \end{pmatrix}$ |
| **Orthorhombic**, $a, b, c$ | $\begin{pmatrix} a^2 & 0 & 0 \\ 0 & b^2 & 0 \\ 0 & 0 & c^2 \end{pmatrix}$ | $\begin{pmatrix} \frac{b\,c}{a} & 0 & 0 \\ 0 & \frac{a\,c}{b} & 0 \\ 0 & 0 & \frac{a\,b}{c} \end{pmatrix}$ | $\begin{pmatrix} 1 & \mathbf{i}_a & \mathbf{i}_b & \mathbf{c} \\ \mathbf{i}_a & -a^2 & \frac{a\,b}{c}\mathbf{i}_c & -\frac{a\,c}{b}\mathbf{i}_b \\ \mathbf{i}_b & -\frac{a\,b}{c}\mathbf{i}_c & -b^2 & \frac{b\,c}{a}\mathbf{i}_a \\ \mathbf{c} & \frac{a\,c}{b}\mathbf{i}_b & -\frac{b\,c}{a}\mathbf{i}_a & -c^2 \end{pmatrix}$ |
| **Hexagonal**, $a, c$ | $\begin{pmatrix} a^2 & -\frac{a^2}{2} & 0 \\ -\frac{a^2}{2} & a^2 & 0 \\ 0 & 0 & c^2 \end{pmatrix}$ | $\begin{pmatrix} \frac{2c}{\sqrt{3}} & \frac{c}{\sqrt{3}} & 0 \\ \frac{c}{\sqrt{3}} & \frac{2c}{\sqrt{3}} & 0 \\ 0 & 0 & \frac{\sqrt{3}a^2}{2c} \end{pmatrix}$ | $\begin{pmatrix} 1 & \mathbf{i}_a & \mathbf{i}_b & \mathbf{i}_c \\ \mathbf{i}_a & -a^2 & \frac{a^2\,(\sqrt{3}\,\mathbf{i}_c + c)}{2c} & -\frac{c\,(\mathbf{i}_a + 2\,\mathbf{i}_b)}{\sqrt{3}} \\ \mathbf{i}_b & \frac{a^2\,(-\sqrt{3}\,\mathbf{i}_c + c)}{2c} & -a^2 & \frac{c\,(2\,\mathbf{i}_a + \mathbf{i}_b)}{\sqrt{3}} \\ \mathbf{i}_c & \frac{c\,(\mathbf{i}_a + 2\,\mathbf{i}_b)}{\sqrt{3}} & -\frac{c\,(2\,\mathbf{i}_a + \mathbf{i}_b)}{\sqrt{3}} & -c^2 \end{pmatrix}$ |
| **Monoclinic**, $a, b, c, \beta$ | $\begin{pmatrix} a^2 & 0 & a\,c\cos\beta \\ 0 & b^2 & 0 \\ a\,c\cos\beta & 0 & c^2 \end{pmatrix}$ | $\begin{pmatrix} \frac{b\,c}{a\sin\beta} & 0 & -\frac{b}{\tan\beta} \\ 0 & \frac{a\,c\sin\beta}{b} & 0 \\ -\frac{b}{\tan\beta} & 0 & \frac{a\,b}{c\sin\beta} \end{pmatrix}$ | $\begin{pmatrix} 1 & \mathbf{i}_a & \mathbf{i}_b & \mathbf{i}_c \\ \mathbf{i}_a & -a^2 & -\frac{b}{\tan\beta}\,\mathbf{a} + \frac{ab}{c\sin\beta}\,\mathbf{i}_c & -a\,c\left(\frac{\sin\beta}{b}\,\mathbf{i}_b + \cos\beta\right) \\ \mathbf{i}_b & \frac{b}{\tan\beta}\,\mathbf{i}_a - \frac{ab}{c\sin\beta}\,\mathbf{i}_c & -b^2 & \frac{b\,c}{a\sin\beta}\,\mathbf{i}_a - \frac{b}{\tan\beta}\,\mathbf{i}_c \\ \mathbf{i}_c & a\,c\left(\frac{\sin\beta}{b}\mathbf{i}_b - \cos\beta\right) & -\frac{b\,c}{a\sin\beta}\,\mathbf{i}_a + \frac{b}{\tan\beta}\,\mathbf{i}_c & -c^2 \end{pmatrix}$ |

For the triclinic structure, the formulae are too long to fit inside the table; they are thus given below

**Triclinic**, $a, b, c, \alpha, \beta, \gamma$

$$\boldsymbol{\mathcal{M}} = \begin{pmatrix} a^2 & a\,b\cos\gamma & a\,c\cos\beta \\ a\,b\cos\gamma & b^2 & b\,c\cos\alpha \\ a\,c\cos\beta & b\,c\cos\alpha & c^2 \end{pmatrix}, \qquad \boldsymbol{\mathcal{X}} = \frac{1}{V'}\begin{pmatrix} \frac{b\,c}{a}(\sin\alpha)^2 & c\,(\cos\alpha\cos\beta - \cos\gamma) & b\,(-\cos\beta + \cos\alpha\cos\gamma) \\ c\,(\cos\alpha\cos\beta - \cos\gamma) & \frac{a\,c}{b}(\sin\beta)^2 & a(-\cos\alpha + \cos\beta\cos\gamma) \\ b\,(-\cos\beta + \cos\alpha\cos\gamma) & a\,(-\cos\alpha + \cos\beta\cos\gamma) & \frac{a\,b}{c}\,(\sin\gamma)^2 \end{pmatrix}$$

and $\boldsymbol{Q}_\mathrm{c} = \begin{pmatrix} 1 & \mathbf{i}_a & \mathbf{i}_b & \mathbf{i}_c \\ \mathbf{i}_a & -a^2 & \boldsymbol{Q}_\mathrm{c}[2,3] & \boldsymbol{Q}_\mathrm{c}[2,4] \\ \mathbf{i}_b & \boldsymbol{Q}_\mathrm{c}[3,2] & -b^2 & \boldsymbol{Q}_\mathrm{c}[3,4] \\ \mathbf{i}_c & \boldsymbol{Q}_\mathrm{c}[4,2] & \boldsymbol{Q}_\mathrm{c}[4,3] & -c^2 \end{pmatrix}$

with

$$\boldsymbol{Q}_\mathrm{c}[3,2] = \frac{b\ c\ (\cos\beta - \cos\alpha\ \cos\gamma)\ \mathbf{i}_a + a\ c\ (\cos\alpha - \cos\beta\ \cos\gamma)\ \mathbf{i}_b - a\ b\ (\sin\gamma)^2\ \mathbf{i}_c}{c\ \mathrm{V}'} - a\ b\ \cos\gamma$$

$$\boldsymbol{Q}_\mathrm{c}[2,3] = \frac{b\ c\ (\ -\cos\beta + \cos\alpha\ cos\gamma)\ \mathbf{i}_a + \ a\ c\ (-\cos\alpha\ + \cos\beta\ \cos\gamma)\ \mathbf{i}_b + a\ b\ (\sin\gamma)^2\ \mathbf{i}_c}{c\ \mathrm{V}'} - a\ b\ \cos\gamma$$

$$\boldsymbol{Q}_\mathrm{c}[4,2] = \frac{a\ b\ (-\cos\alpha + \cos\beta\ \cos\gamma)\ \mathbf{i}_c + b\ c\ (-\cos\gamma + \cos\alpha\ \cos\beta)\ \mathbf{i}_a + a\ c\ (\sin\beta)^2\ \mathbf{i}_b}{b\ \mathrm{V}'} - a\ c\ \cos\beta$$

$$\boldsymbol{Q}_\mathrm{c}[2,4] = \frac{a\ b\ (\cos\alpha - \cos\beta\ \cos\gamma)\ \mathbf{i}_c + b\ c\ (\cos\gamma - \cos\alpha\ \cos\beta)\ \mathbf{i}_a - a\ c\ (\sin\beta)^2\ \mathbf{i}_b}{b\ \mathrm{V}'} - a\ c\ \cos\beta$$

$$\boldsymbol{Q}_\mathrm{c}[4,3] = \frac{a\ c\ (\cos\gamma - \cos\alpha\ \cos\beta)\ \mathbf{i}_b + \ a\ b\ (\cos\beta - \cos\alpha\ \cos\gamma)\ \mathbf{i}_c - b\ c\ (\sin\alpha)^2\ \mathbf{i}_a}{a\ \mathrm{V}'} - b\ c\ \cos\alpha$$

$$\boldsymbol{Q}_\mathrm{c}[3,4] = \frac{a\ c\ (-\cos\gamma + \cos\alpha\ \cos\beta)\ \mathbf{i}_b + \ a\ b\ (-\cos\beta + \cos\alpha\ \cos\gamma)\ \mathbf{i}_c + b\ c\ (\sin\alpha)^2\ \mathbf{i}_a}{a\ \mathrm{V}'} - b\ c\ \cos\alpha$$

with $V' = \frac{\sqrt{det\boldsymbol{M}}}{a\ b\ c} = \sqrt{1 - (\cos\alpha)^2 - (\cos\beta)^2 - (\cos\gamma)^2 + 2\ \cos\alpha\ \cos\beta\ \cos\gamma}$